\documentclass[preprint,authoryear,review,12pt]{elsarticle}

\usepackage{graphicx}
\usepackage{amssymb}
\usepackage{subfigure}
\usepackage[pagewise, modulo]{lineno} 
\usepackage{graphicx}
\usepackage{ifthen}
\usepackage{epstopdf}
\usepackage{aastex_hack}
\usepackage{fancyvrb}
\usepackage{amsmath}
\fvset{baselinestretch=1}
\usepackage{acronym}

\usepackage{pdfpages}

\newcommand{\cf}{cf.~}

\newcommand{\eg}{e.g.,~}

\newcommand{\ie}{i.e.,~}

\newcommand{\vs}{vs.\ }

\newcommand{\sect}[1]{Sec.~\ref{s:#1}}

\newcommand{\fig}[1]{Fig.~\ref{f:#1}}
\newcommand{\figtwo}[2]{Figs.~\ref{f:#1} and \ref{f:#2}}
\newcommand{\figs}[2]{Figs.~\ref{f:#1}--\ref{f:#2}}
\newcommand{\Fig}[1]{Figure~\ref{f:#1}}

\newcommand{\tbl}[1]{Table~\ref{t:#1}}

\newcommand{\code}[1]{\texttt{#1}}

\journal{Planetary and Space Science}

\begin{document}

\begin{frontmatter}



\title{Low-speed Impact Simulations into Regolith in Support of Asteroid Sampling
  Mechanism Design I.: Comparison with 1-g Experiments}

\author[label1,label2]{Stephen~R.~Schwartz\corref{cor}}
\ead{srs@oca.eu}

\author[label2]{Patrick~Michel}
\ead{michelp@oca.eu}

\author[label1]{Derek~C.~Richardson}
\ead{dcr@astro.umd.edu}

\author[label3]{Hajime~Yano}
\ead{yano.hajime@jaxa.jp}

\cortext[cor]{\\Corresponding author (Phone/Fax): +33 4 92 00 30 (55/58)}

\address[label1]{
  Department of Astronomy, University of Maryland, College Park MD 20740-2421\\
  \bigskip
}
\address[label2]{
  Lagrange Laboratory, University of Nice Sophia Antipolis, CNRS\\
  Observatoire de la C\^ote d'Azur, C.S. 34229, 06304 Nice Cedex 4, France\\
  \bigskip
  }
\address[label3]{
Department of Interdisciplinary Space Science, Institute of Space and Astronautical Science,
  Japan Aerospace Exploration Agency, 3-1-1 Yoshinodai, Chuo-ku, Sagamihara, Kanagawa 252-5210, JAPAN\\
}


\begin{abstract}

This study is carried out in the framework of sample-return missions to asteroids that use a
low-speed projectile as the primary component of its sampling mechanism (\eg JAXA's
Hayabusa and Hayabusa2 missions).
We perform numerical simulations of such impacts into granular materials
using different projectile shapes under Earth's gravity.  We then compare the amounts of
ejected mass obtained in our simulations against what was found in experiments that used
similar setups, which allows us to validate our numerical approach.  We then investigate the
sensitivity of various parameters involved in the contacts between grains on the amount
of mass that is ejected.
For the targets, we consider 2 different monodisperse grain-diameter sizes: 5~mm and 3~mm.  The impact speed of the projectile is 11~m~s$^{-1}$, and is directed downward,
perpendicular to the surface of the targets.  Using
an implementation of the soft-sphere discrete element method (SSDEM) in the $N$-Body gravity tree code \code{pkdgrav}, previously validated in the context of low-speed impacts into
sintered glass bead agglomerates, we find a noticeable dependence of the amount of ejected mass on the projectile shape.  As found in experiments, in the case of the larger target grain size (5~mm), a conically shaped projectile
ejects a greater amount of mass than do projectiles of other shapes, including disks and spheres.
We then find that numerically the results are sensitive to the normal
coefficient of restitution of the grains, especially for impacts into targets comprised of smaller
grains (3~mm).  We also find that static friction plays a more important role for impacts into
targets comprised of the larger grains.  As a preliminary demonstration, one of these
considered setups is simulated in a microgravity environment.  As expected, a reduction in
gravity increases both the amount of ejected mass and the timescale
of the impact process.  A dedicated quantitative study in microgravity is the subject of future
work.  We also plan to study other aspects of the ejection process such as velocity distributions
and crater properties, and to adapt our methodology to the conditions of sampling mechanisms
included in specific mission designs.

\end{abstract}


\end{frontmatter}


\section{Introduction}
\label{s:int4}

The impact process plays a major
role in the formation and evolution of planetary systems,
including our own Solar System.  It is particularly important because impact
craters are the most commonly observed geological features
on the surfaces of solid Solar System
bodies.  Crater shapes and features are crucial
sources of information regarding past and present surface
environments, and can provide us indirect information
about the internal structures of these bodies as
well.  Piecing together the chronology of these surfaces relies on
our ability to measure the size distribution of craters and to
discriminate between primaries and secondaries, the
latter the result of low-speed impact ejecta falling back upon
the surface.
Low-speed impact mechanisms are also being purposefully
implemented into the design of robotic spacecraft missions
as a way to collect samples from the surfaces of small bodies
such as asteroids and comet nuclei.   Specific mechanisms have
been incorporated, for instance, by JAXA's Hayabusa and
Hayabusa2 missions \citep{Fujiwara2006,Tachibana2013}.
The Hayabusa2 mission is targeted for launch in late 2014, and
aims to collect surface and sub-surface samples from the primitive
near-Earth (C-type) asteroid 1999JU$_3$ \citep{Vilas2008} by firing
a small, low-speed, semispherical projectile, similar to the projectile
aboard the previous Hayabusa mission
\citep{Fujiwara2005,Yano2006}.  The
Hayabusa2 sample is scheduled to be returned to Earth in 2020.

In this paper we investigate the influence of various projectile
shapes, for use in performing low-speed impacts, on the
amount of ejected mass from coarse granular material targets.  This
allows for the determination of the optimal
shape for maximizing the amount of material ejected (and thus
collected) from surfaces composed of granular material for purposes of
collector design aboard sample-return missions.  Moreover, the
outcomes are also relevant to the study of secondary cratering,
the result of re-impaction of ejecta following a larger impact event,
since the outcomes of such processes may be sensitive to
the projectile shape.

The rationale in the assumption that asteroid surfaces consist
of granular material is based on the results of several observations.
First confirmed by space missions that have visited
asteroids in the last few decades \citep{Veverka2000,Fujiwara2006},
it appears that all encountered asteroids thus far are covered with
some sort of granular material, usually referred to as ``regolith."
To date, this includes a large range in asteroid sizes, from the largest one visited,
by the Dawn spacecraft, the main belt asteroid (4) Vesta, which measures
${\sim}500$~km across, to the smallest
one, sampled by the Hayabusa mission, the NEA (25143) Itokawa, which
measures ${\sim}500$~m across \citep{Russel2012,Yano2006,Miyamoto2007}.
Thermal infrared
observations support the idea that most asteroids are covered with
regolith, given their preferentially low thermal inertia
\citep{Delbo2007}.  Thermal inertia measurements also point to a
trend based on asteroid size:
larger objects are expected to have a surface covered by a layer of
fine regolith, while smaller ones are expected to have a surface
covered by a layer of coarse regolith \citep{Delbo2007,Muller2013T}.

\begin{figure*}  
  \centering
  \epsscale{1}
  \raggedright
  \plotone{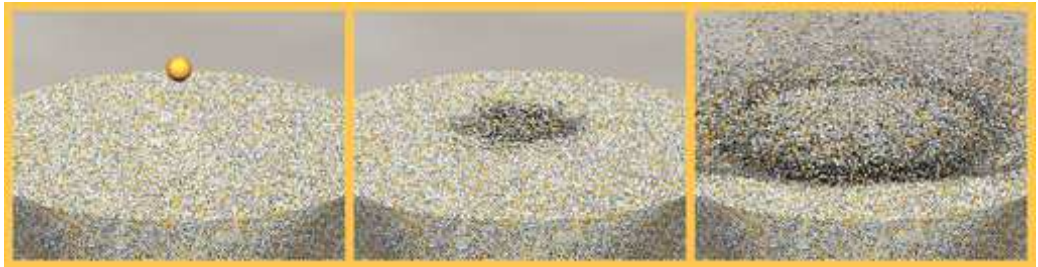}
  \caption[Cratering simulation: 100~m/s projectile into 1,137,576~particles]{
  Cratering simulation using a target comprised of 1,137,576~particles.  A 9~cm-radius
  projectile impacts
  perpendicular to the surface at a speed of 100~m/s into a 155~cm-radius half-shell filled with
  1~cm-radius grains of collisional restitution coefficient 0.2.  From left to right: 5~ms prior
  to impact; 15~ms after impact; and 375~ms after impact.
  }
  \label{f:light2}
\end{figure*}

In previous studies, we have satisfactorily
reproduced the experimental results of
low-speed impacts into sintered glass bead agglomerates
\linebreak\citep{Schwartz2013}.  We have also demonstrated the ability to
simulate the evolution of millions of granular particles
in the context of both flow from a granular hopper 
\citep{Schwartz2012} and low-speed cratering events
(\eg \citealt{Schwartz2012DPS}); the latter included an
evaluation of ejecta speeds and trajectories, and a
preliminary analysis of resulting crater sizes and morphologies
at the site of the impact (see \fig{light2}).

In principle, if the cratering process involves monolithic
rock and/or if the impact speed is in the hypervelocity regime
(\ie is larger than the sound
speed of the material), then hydrocode simulations
that take into account large plastic deformations and phase
changes of particles are the most adapted to model the
process \citep{Barr2011}.  However, if the cratering process
involves
a low-speed impactor into granular material, then the
discreteness of particles as well as the different contact
frictional forces among them must be taken into account.
Sophisticated constitutive equations may be
implemented into hydrocodes to study these types of
cases, but numerical codes capable of directly simulating
the evolution of particles and the contact forces
between them during such impact events are probably
best suited.

We use an implementation of the Soft-Sphere
Discrete Element Method (SSDEM), as developed in
\citet{Schwartz2012}, to model the impact cratering process
into granular materials to predict the amount of ejected mass.
The numerical study presented here is based on the experimental
results found by \citet{Makabe2008Yano} in preparation for the Haybusa2
mission, investigating the effects of different projectile shapes on
the amounts of collected mass.  We then perform simulations of their
experiments, to compare the outcomes and provide results for a wider
parameter space.
Throughout this paper, when particle or projectile sizes are given, they can be assumed
to represent their diameters, unless explicitly stated otherwise.

In the current study, our primary aim is to focus on the same measured outcomes from the
\cite{Makabe2008Yano} experiments: the amount of ejected mass.  We leave for future study the investigation of
other interesting aspects of low-speed impacts (\eg crater size, morphology, etc.).  For this initial study, certain adaptations to the
numerical code had to be made in order to simulate projectiles with shapes other than spheres
(see \sect{method4}).  As will be explained, the shapes of the projectiles need to be modeled explicitly.

Many laboratory studies analyzing different aspects of the cratering process caused by
low-speed (sub-sonic) impacts into loose granular material have already been performed.
These include studies into the properties of crater growth and ejecta fate, and their
correlations
to impact conditions such as the impact speed, target material, projectile material, and
the gravitational environment
(see, \eg \citet{Uehara2003}, \citet{Yamamoto2006}, \citet{Yamamoto2009}, who consider
crater depth and morphology; \citet{Yamamoto2005}, \citet{Housen1983}, who investigate
particle ejecta speeds; \citet{Nakamura2013}, who consider the reaction force on the
projectile as the impact process progresses; and \citet{Wada2006}, for a numerical study of the excavation stage of low-speed impacts into regolith).

In \sect{exp4}, we give an overview of the experiments performed by \linebreak\citet{Makabe2008Yano}.  We then
describe our numerical method used to perform the simulations of the low-speed impacts with
specific projectile shapes in \sect{method4}.  In \sect{sim4}, the simulations are presented, and their results
are given.  And finally, in \sect{con4}, discussions, conclusions, and perspectives are presented.

\section{Laboratory impacts into granular material}
\label{s:exp4}

\begin{figure*}  
  \centering
  \epsscale{0.19}
  \raggedright
  \plotone{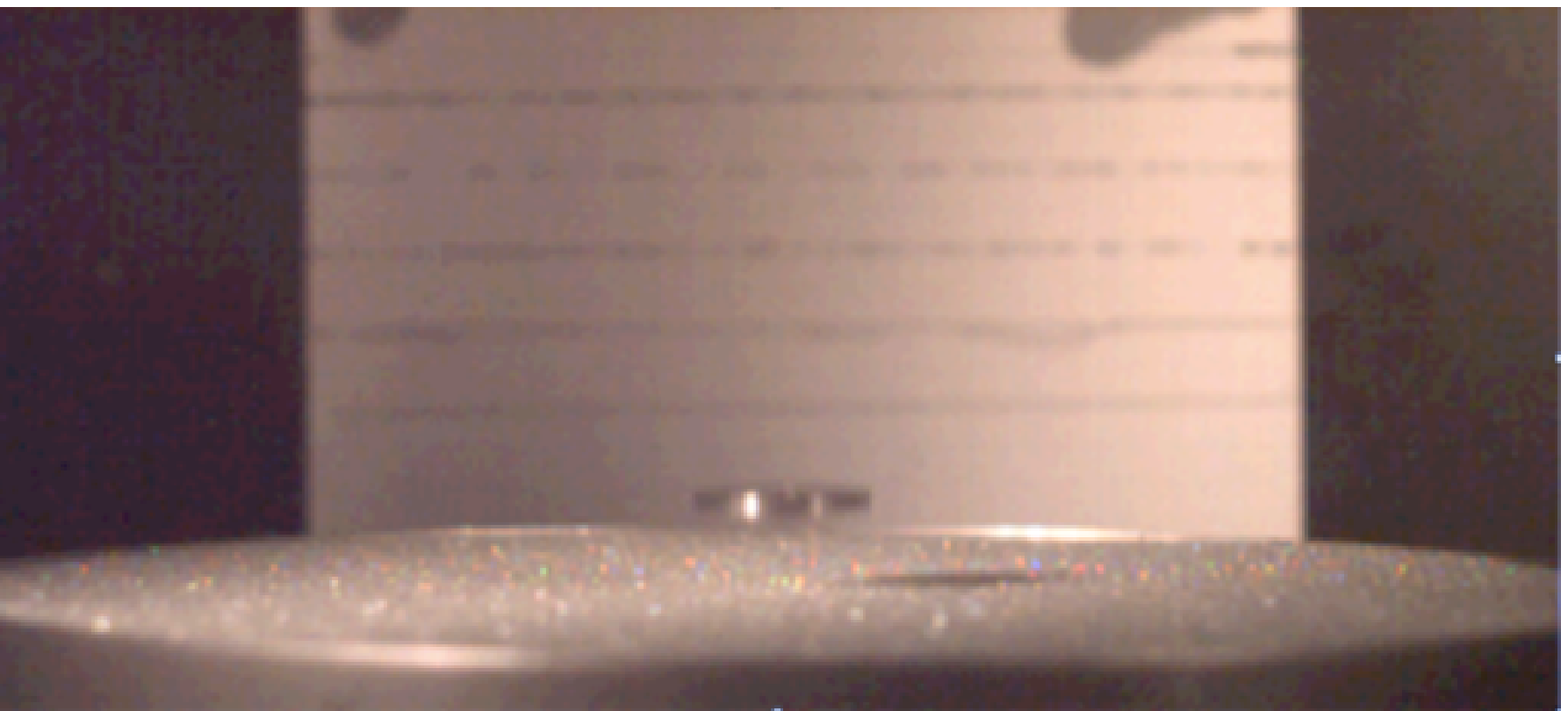}
  \plotone{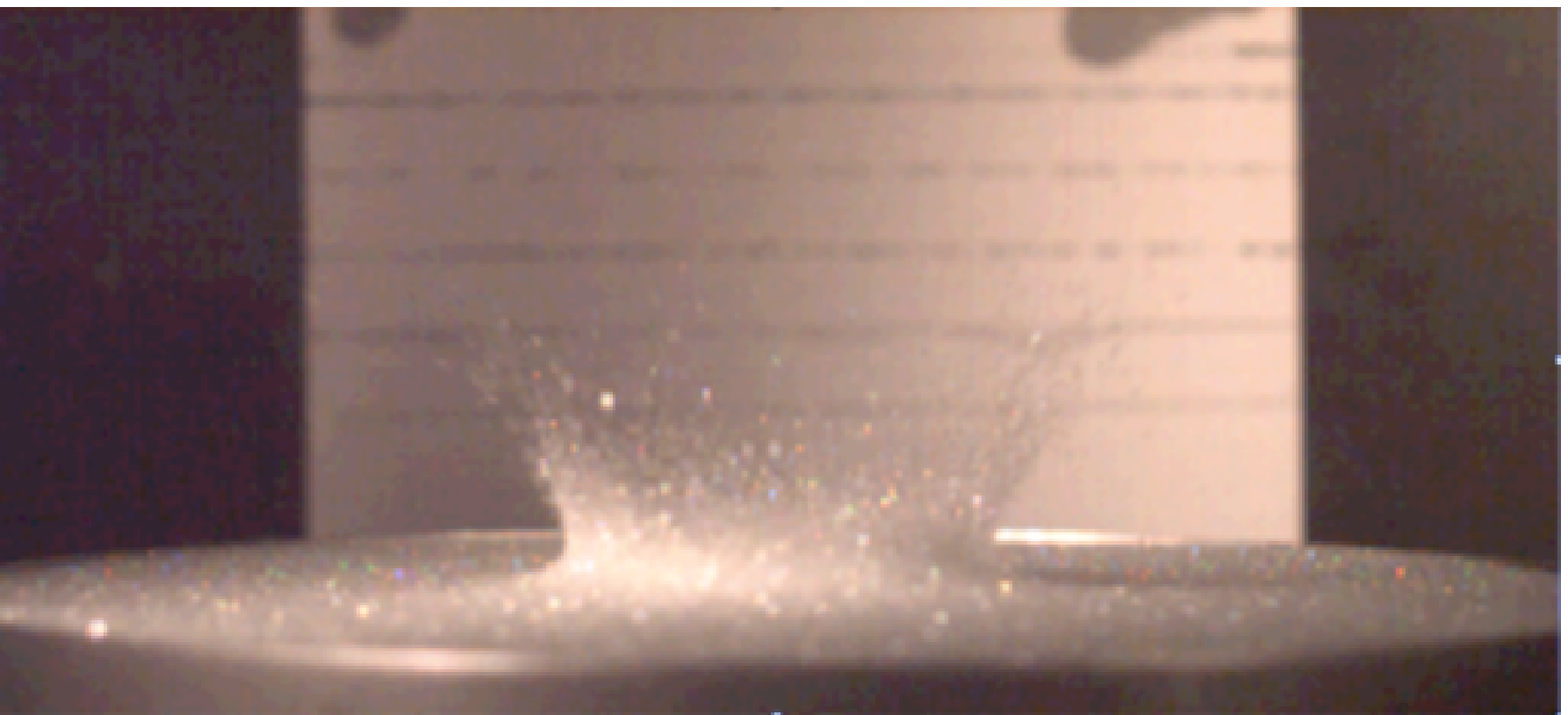}
  \plotone{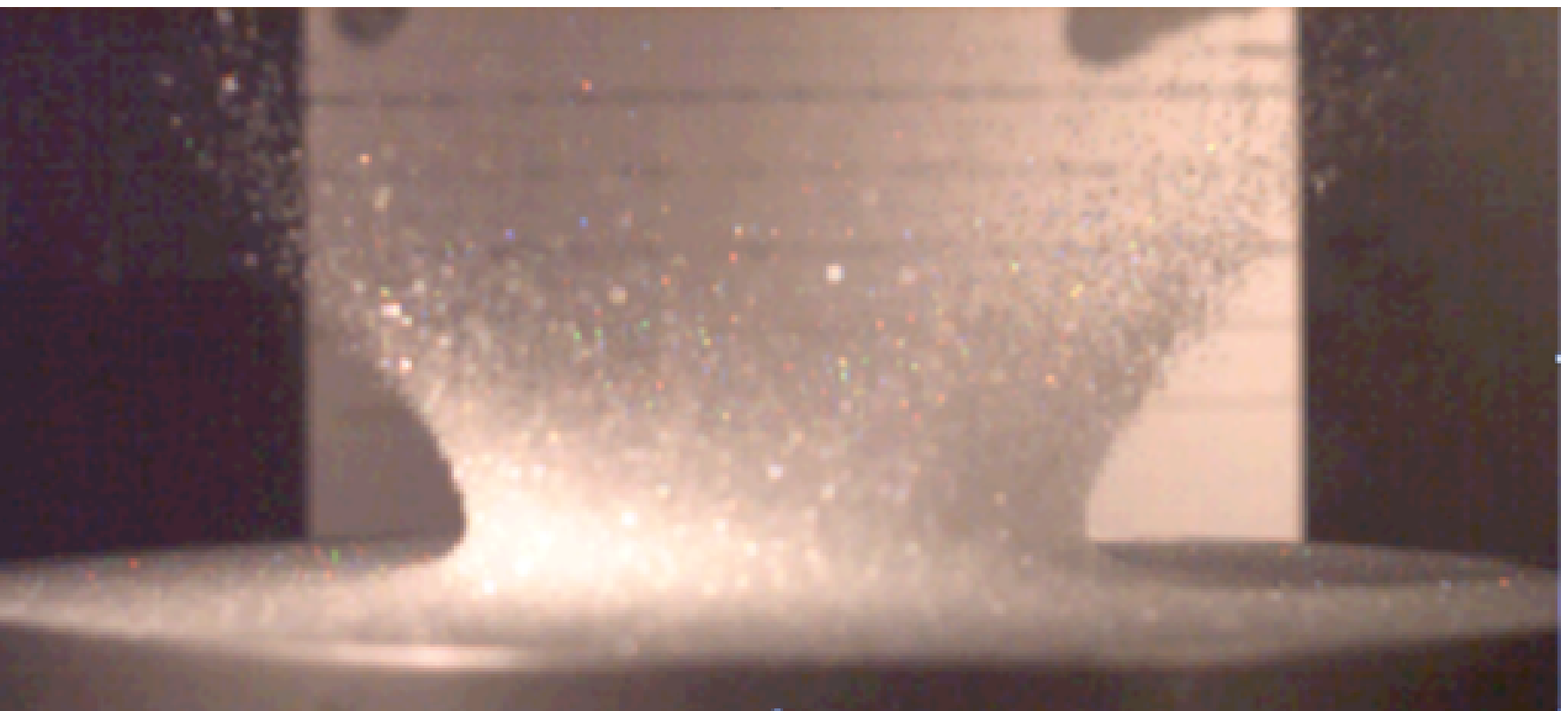}
  \plotone{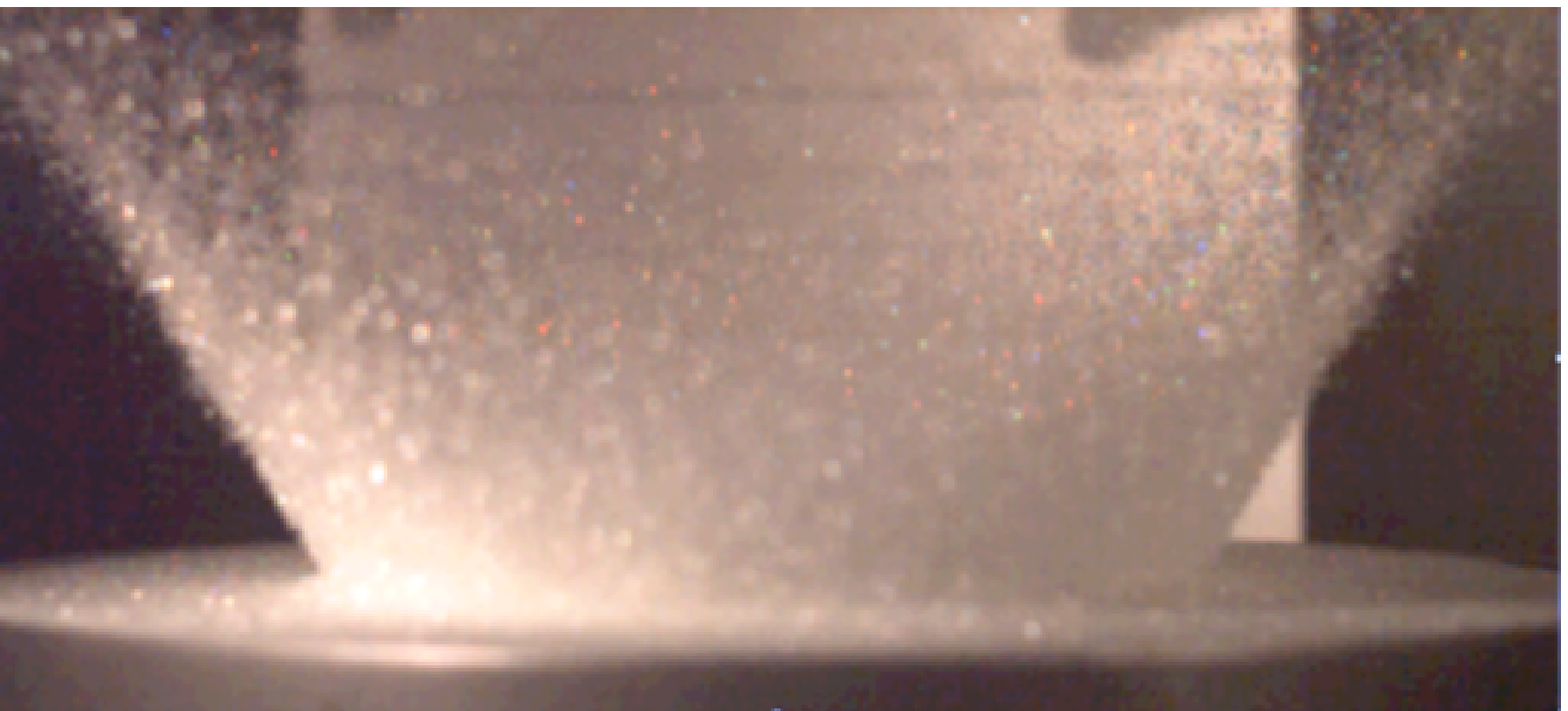}
  \plotone{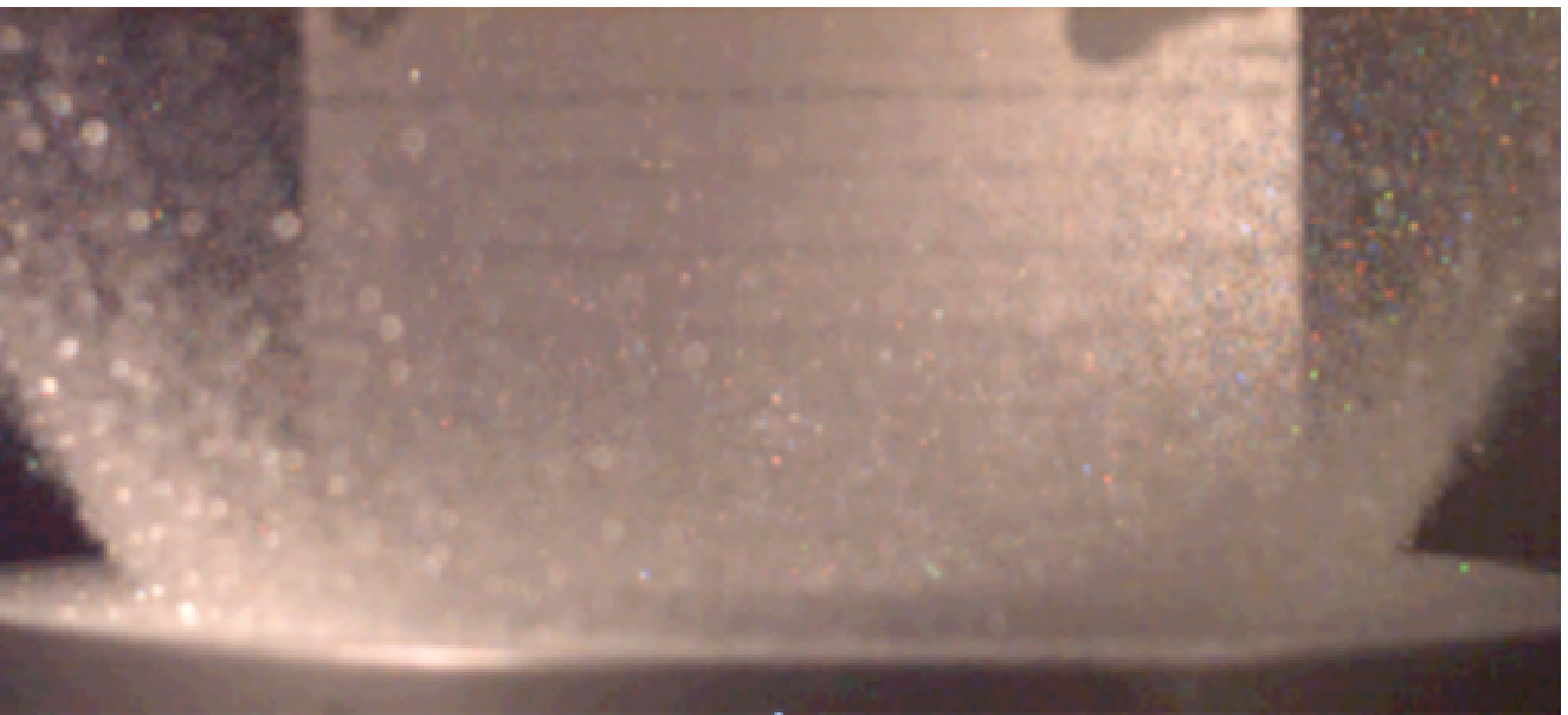}
  \caption[Impact experiment into 0.5~mm glass beads]{
  Successive frames showing an experiment that uses a flat projectile impacting into a target
  comprised of 0.5~mm glass beads, producing a symmetrical ejecta plume.
  }
  \label{f:500m1}
\end{figure*}

\begin{figure*}  
  \centering
  \epsscale{1}
  \raggedright
  \plotone{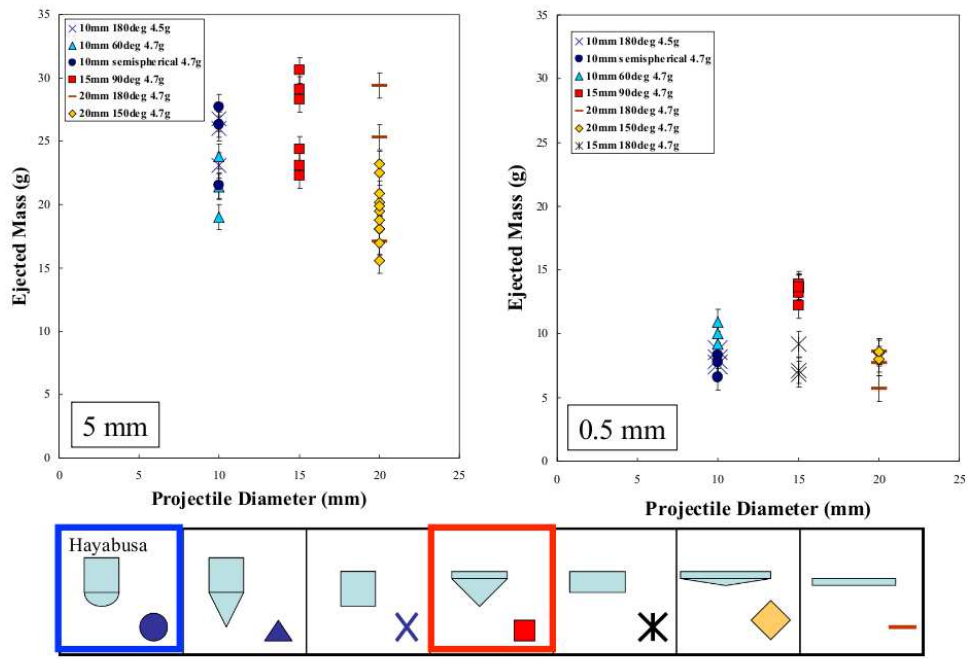}
  \raggedleft
  {\footnotesize Credit: \textit{\citet{Makabe2008Yano}}}
  \caption[Experimental results: ejected mass obtained in laboratory]{
  Total mass ejected using 7 impactor shapes (see labels on the plots).  Left: target comprised
  of 5~mm glass beads.  Right: target comprised of 0.5~mm glass beads.  All projectiles
  were directed downward at 11~m~s$^{-1}$, impacting the surface with a normal
  (perpendicular) impact angle.
  }
  \label{f:exp}
\end{figure*}

\citet{Makabe2008Yano} performed low-speed impacts into containers filled with coarse glass
beads by shooting projectiles of approximately the same mass as the projectiles
designed for the sampling mechanism aboard the Hayabusa spacecraft (\ie ${\sim}5$~g
in mass) but considering a variety of shapes (\fig{500m1}).  The
primary purpose of this experiment was to look for
substantial improvement of this impact sampling mechanism, in particular, for surfaces of
coarse-grained regolith.  A good example of such a surface is the smooth terrain of the
MUSES-C region on Itokawa, which was discovered and characterized by the Hayabusa
mission \citep{Yano2006,Miyamoto2007}.
The portion of the experiment that we consider in this numerical study involves low-speed
impacts, ${\sim}11$~m s$^{-1}$, of 7 different types of projectiles into cylindrical containers, 200~mm in diameter
and 150~mm in height, filled with approximately monodisperse glass beads.  Two different sizes of glass beads were used
for the targets, 5~mm and 0.5~mm, chosen to produce projectile-to-target grain size ratios of
2:1, 3:1, and 4:1 for the targets comprised of larger grains, and 20:1, 30:1, and 40:1 for
the targets comprised of smaller grains.  By comparing the pre-impact weight to the
post-impact weight of the targets, the amount of material ejected outside the
container was measured for each of the impact experiments across the range of different
projectile shapes.  All projectiles were made of SUS304 metal (common steel, 8.00 g~cm$^{-3}$), the 7 projectile shapes were
[mass, diameter (or diameter of the base for cones), angle of projectile surface at impact point]:
a hemisphere, as an analog of the projectile chosen for the mechanism aboard Hayabusa
[4.7~g, 15~mm, rounded/spherical], three flat disks [4.5~g, 10~mm, 180 degrees; 4.7~g,
15~mm,
180 degrees; 4.7~g, 20~mm, 180 degrees], and three cones [4.7~g, 10~mm, 60 degrees;
4.7~g, 15~mm, 90 degrees; 4.7~g, 20~mm, 150 degrees].  The experimental results are shown
in \fig{exp} (see \citealp{Makabe2008Yano} for further details of the methodology and
results).  Of these shapes, it was determined that the 90-degree cone tended to produce the greatest amount of ejected material. 

\section{Numerical method}
\label{s:method4}

We attempt to replicate numerically, the experiment described in \sect{exp4} as a verification
test of the code's applicability in the context of granular impacts and its
usefulness in simulating specific aspects of the sampling mechanism that can be designed
and used on future sample return missions.  We use an implementation of the Soft-Sphere
Discrete Element Method (SSDEM), as developed in
\citet{Schwartz2012} for the $N$-Body code \code{pkdgrav} \citep{Stadel2001},
a numerical gravity solver, adapted for planetary science applications
by \citet{Richardson2000}.  In order to
perform a satisfactory replication of the types of experiments described in \sect{exp4},
we first need support for non-spherical projectile shapes.  Thought was given to the idea of
making these shapes out of rigid aggregates \citep{Richardson2009} or out of cohesive
soft-sphere agglomerates \citep{Schwartz2013}.
To date, the collisional routines in \code{pkdgrav} have been limited by the fact that
particles in the simulation are taken to be spherical.
Spherical particles are convenient because intersections are easy to detect: if the sum
of the radii of two particles is larger than the distance between their centers-of-mass (COMs),
then the two particles are in contact.  Further, the plane of contact between a particle and its
neighbor is easier to locate in the case of two overlapping spheres than it is for objects of
arbitrary shape (\cf \citealt{Schwartz2012}).
This owes to the fact that alignment between spherical bodies is largely irrelevant and that
every point on a spherical particle's (undeformed)
surface is known given the radius and COM location of the particle.  This simplifies the coding
and lessens the number of computations needed to be performed during a given integration
timestep, making the simulation run quickly.
However, in the real world, perfectly spherical grains are difficult to find, even in
a controlled laboratory setting, much less in the surface regolith that appears to cover
the vast majority of
solid bodies in the Solar System.  Since these effects could be important, tools have been
developed for \code{pkdgrav}
that help account for the effects of non-spherical objects.  These include the
use of movable, but non-reactionary, boundary walls
\citep{Richardson2011,Schwartz2012}.  Effects of
non-sphericity are accounted for with the use of ``quasi-reactionary" boundary walls in place
of the particle (described below).
We have also introduced complex wall geometries to account for non-spherical projectiles
involved in our low-speed impact models \citep{Schwartz2012}.

Relative to the Earth's gravitational field, self-gravity between particles is insignificant and
was turned off for these simulations to save computational time.  The neighbor-finding routine
of the tree code, used to quickly and efficiently perform interparticle gravity calculations, however, is still used to speed up neighbor searches.  It generates a list of closest neighbors for each particle at each
timestep in order to check for overlap (contacts); and at every timestep, every particle is
checked for states of overlap with its closest neighbors (without redundancy), the
projectile, and each boundary wall.  Further details, including the specific treatments of
the different types of particle-wall interactions, are contained in \citet{Schwartz2012}.

For simulations under Earth's gravity conditions, two targets were prepared, one using 5-mm
monodisperse particles and one using 3-mm
monodisperse particles.  As stated in \sect{exp4}, the laboratory experiments involved targets comprised
of 5-mm particles and 0.5-mm particles.  Although it is feasible to calculate an impact on a
target comprised of several tens of millions of particles, which is what would be required for
the 0.5-mm case, we chose to use particles of 5~mm and 3~mm in order to perform a
number of simulations in a reasonable amount of time, and to nonetheless
attempt to assess the effects of grain size on the outcome.

To prepare a given target, we suspended arrays of particles above an empty cylindrical
container, the same size used in the experiments (described in \sect{exp4}), and then simulated a funnel that allowed these particles to empty into the container
gradually under the influence of Earth's gravity.  In order to further randomize the granular configuration, we subjected the container to forced oscillations along the axis of the cylinder, jostling the particles, which we then allowed to resettle inside.  We then deleted those particles located above the rim of the container, causing in turn a very slight readjustment of those particles remaining inside the container (since the configurations of soft-sphere particles are sensitive to changes in confining pressure).  After ensuring that all motion had once again seized, the
targets ended up with packing densities of 62\% and 63\% for the 5-mm-bead
targets and the 3-mm-bead targets, respectively, comparable to the targets in the laboratory, which had estimated packing densities of around 64\%.

We measured the amount of ejected mass by the same criterion used experimentally, namely, by counting a grain as having been ejected if and only if it escapes the container (particles lofted as a result of the impact, but which land and remain inside the container do not count toward the ejected mass, since they would not have been counted in the laboratory experiment).  As each impact simulation is running, periodic checks are made to analyze the position and momentum-state of each grain.  Grains below the rim of the container with COM distances from the vertical axis greater than the radius of the container are considered ejected.  For each grain above the rim of the container, the code outputs its position, its vertical and radial velocity components, and, assuming a parabolic (collisionless) trajectory, the time when and position where it will cross the plane of the container top.  When grain centers are located no higher than about 3 grain diameters above the rim and are effectively still (speeds $\lesssim~10^{-3}$ cm~s$^{-1}$), the simulation is assessed qualitatively using the ray-tracing visualization tool POV-Ray to build an animation.  If we confirm both quantitatively and visually that further integration would not yield the ejection of additional grains from the container, then the mass located at distances from the vertical axis greater than the container radius are summed; this sum is taken to be the ejected mass. 

Since walls are normally not configured to feel any reaction force, in order to
correctly account for resistance
by the granular target to penetration, the projectiles formed by combining wall primitives were
given special treatment.  For simplicity, and taking advantage of the cylindrical symmetry in
each of the impacts described in \sect{exp4}, only the motion along the track of the
projectile's initial trajectory is affected.  The
implicit assumption is that momentum transfer from the granular material onto the
projectile's COM, along the direction of its initial trajectory, should account for the
majority of the total work that the material does on the projectile in the actual experiment.

\begin{figure*}  
  \centering
  \epsscale{1}
  \raggedright
  \plotone{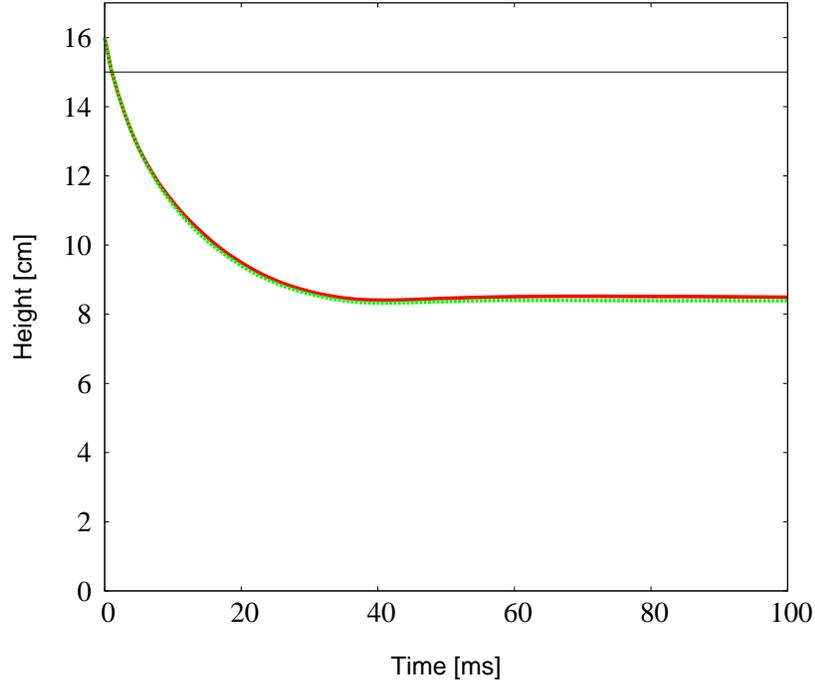}
  \caption[Projectile penetration depths: free soft-sphere particle {\vs}shell]{
  Height of projectile as a function of time using different representations of a
  spherical intruder for two otherwise identical simulations.  The solid (red) line represents
  the free soft-sphere particle, the dashed (green) line represents the shell of equal size
  and mass, and the thin solid line at 15~cm shows the initial surface height of the
  target.  Outcomes of the two cases are similar, with the final depth of the projectiles differing
  by less
  than a single target particle radius, justifying our methodology in situations where the effects
  from torques and lateral forces on the projectile can be ignored safely.  Also evident is a
  slight rebound effect of the projectiles after achieving their maximum depths (they then exhibit
  highly damped oscillation around their final, equilibrated penetration depths).
  }
  \label{f:penetrate}
\end{figure*}

The acceleration of an individual particle (grain) due to its collision with the impactor
relative to
the impactor is given by $F/\mu$, where $F$ is the force exerted on the particle and $\mu$
is the reduced mass of the grain-projectile system.
In these simulations, since the projectile mass is much larger than the mass of individual
grains in the considered cases, we take $\mu$ to be equal to the
mass of the grain for grain-projectile collisions.  For instance, the mass of an individual
5-mm glass bead is 0.163~g, and the mass of each projectile, which matches its
respective counterpart in the laboratory, is 4.7~g (4.5~g, in the case
of the flat 10-mm disk)---this gives a reduced mass of 0.158~g; for the 3-mm glass bead
targets, the mass and reduced
mass of the target particles are 0.0353~g and 0.0351~g, respectively.  This slight violation of the law of
conservation of energy is not a
concern in this scenario since there are significant damping effects (the coefficient of
restitution of the grains is less than
unity) that utterly dominate the energy loss to the system.  \Fig{penetrate} shows a
comparison of the penetration depth of the projectile as a function of time from two
different impact
simulations into targets comprised of 3-mm glass beads: one simulation uses a
projectile made from a soft-sphere particle, and the other uses a projectile made of a ``shell"
wall primitive, both equal
to the size and mass of the semispherical projectile used in the experiments (note that the
hemisphere used in the experiments was substituted for a sphere in the simulations, but both
show the same surface to the target upon impact).  It can be seen that the path of
the free soft-sphere projectile particle is similar to the path followed by the shell projectile.
Three factors contribute to the slight differences in the curves: 1) the assumption that
$\mu=m$, as discussed; 2) the fact that torques (although the projectile is a sphere, it can still
receive a torque due to surface friction effects) and translational motion parallel to the surface that are
imparted upon the soft-sphere projectile particle are ignored in the case of the shell projectile;
3) differences in numerical roundoff error as the order of mathematical operations may be
different (\eg ``${\rm a}+{\rm b}={\rm b}+{\rm a}$" does not
hold for finite-precision floating-point operations).  However, the fact that the latter option might even be a
consideration points to the similarity
of these two simulations (the final penetration depths of the soft-sphere projectile and shell
projectile are 66.22 and 65.31~mm, respectively, a difference of less than 1~mm or about 1 part
in 72) and shows that the wall primitive serves
as a good proxy for a soft-sphere particle impactor.  For different shapes, the second factor above could become
more of a concern; this is discussed further in \sect{con4}.
Any projectile shape that can be created using arbitrary numbers and combinations of the six wall primitives
described in \citet{Schwartz2012} can be substituted for the shell.
In this study, disks and cones of infinitesimal thickness are also used to simulate projectile shapes.

\section{Numerical simulations and results}
\label{s:sim4}

In order to develop good numerical representations of these impact experiments, parameters
of the target material must first be constrained.  We chose values for the normal and
tangential soft-sphere stiffness parameters to $k_n=1.37$~kg~s$^{-2}$ and $k_t = (2/7)k_n$,
respectively, such that the simulated grains are sufficiently stiff to suffer overlaps of only
1--2\% of their radii even in the case of a direct hit from a 11~m~s$^{-1}$ projectile (as a
utility that accompanies the collision code, additional software has been written to
compute appropriate values of the stiffness parameters).  In performing this study, the same
values of the stiffness parameters, $k_n$ and $k_t$, were used to govern every collision,
\ie grain-grain, grain-projectile, and grain-container.

Holding these stiffness parameters fixed, there are 5 more free parameters in
these simulations that correspond to some of the physical properties of the granular material
and that are also not constrained by the experiment: the coefficient of static friction $\mu_s$,
the coefficient of rolling friction $\mu_r$, the coefficient of twisting friction $\mu_t$, the
normal restitution coefficient $\varepsilon_n$, and a measure of the kinetic damping effects in
the tangential direction $\epsilon_t$ (see, \eg \citealt{Schwartz2012} for detailed definitions).  The
projectile also requires specification of these same free parameters.

For simplicity, we prescribed that a grain in contact with either the projectile or the
container is governed by conservative values of $\varepsilon_n$ and $\epsilon_t$ of 0.5 and
1.0, respectively, \ie moderate collisional dissipation and no sliding friction.  Other material friction
parameters for grain-projectile and grain-wall contacts, $\mu_s$, $\mu_r$, and $\mu_t$, were
taken to be identical to the grain-grain contacts of the particular simulation (see below).  Since
only a relatively small fraction of particles interact directly with either the projectile or the container, these values were not tuned to match the experiments; instead, plausible values were chosen.

For the targets, we used 2 grain
configurations, one corresponding to simulations that involved impacts into 5-mm grains and the other corresponding to simulations that involved impacts into 3-mm grains.  We picked one spot within
a 2-mm square region at the center of the targets' surfaces, at random, as the primary
impact point for every projectile; this point lies 1.14 mm from the center.  This was done in part to control for variations in the precise
bead configurations and impact points over the range of parameter space investigated,
allowing us to measure the effects of each parameter using a limited number of simulations.
As an additional simulation suite, we also chose 2 impact scenarios for which we varied the point of impact, one involving the
5-mm-grain target, and one involving the 3-mm-grain target, in order to assess how a change
in the precise impact point may affect the amount of material ejected (discussed below in this
section).

To sweep this parameter space and analyze the dependencies of these parameters
($\mu_s$, $\varepsilon_n$, $\epsilon_t$, $\mu_r$, and $\mu_t$) on the amount of material
ejected, we performed a large suite of simulations using one experimental configuration
(the same shape impactor and target grain size), varying the values of these 5 parameters.
The experimental configuration that we chose to use as this baseline configuration was the
collision involving the 90-degree cone projectile into the target comprised of 5-mm
grains.  This projectile was chosen because it was highlighted in the experiments as
being more efficient than a semispherical projectile at ejecting mass from a target
\citep{Makabe2008Yano}.  The grain size of 5~mm was chosen because
this matched an actual experiment to simulate coarse grains in the size order that the
Hayabusa spacecraft (spatial resolution of 6--8~mm-per-pixel) observed in the gravel field of
the MUSES-C region on Itokawa.  We thus performed 93 impacts into the 5-mm-bead targets
using the 90-degree
solid-angle-conical impactor.  These 93 impacts each used different values of the five
material parameters mentioned.  The latter 2
parameters, $\mu_r$ and $\mu_t$, were found to have less influence on the amount of
material ejected (see the discussion below).
From these 93 parameter sets, we chose 5 sets (PS$_1$--PS$_5$; see \tbl{impact_params})
that each matched the mass-loss results from the baseline experiment, and that also
exhibited variety in the parameter values.  We then used these 5
sets of parameters with each of the 7 impactor shapes, and each of the 2 targets,
comprised of monodisperse 5-mm grains and monodisperse 3-mm grains, giving a total of
70 impact simulations.  Using one of these parameter sets (PS$_2$),
the 90-degree cone impactor, and the 2 targets (5-mm grains and 3-mm grains), we
performed 16 additional impacts (8 on each target), varying the precise point on the target
bed where the projectile collides to help assess the importance of the precise point of impact (the result of the impact may differ between head-on collisions and collisions between grains).  The 8 additional impact points were distributed uniformly on a circle of radius 2~mm centered on the original impact point.  For 5-mm grains, over the sample of the 9 impacts (primary impact point plus the 8 additional impacts), the measured amount of ejected mass was 28.0343 +/- 1.9422~g (1-$\sigma$ deviation).  The scatter is even less in the case of 3-mm grains, where the measured amount of ejected mass was 29.3830 +/- 0.9353~g.

In all, 179 impacts were simulated using multi-core parallel processing; each
simulation ran for between one day and two weeks, depending on the target and the
processor speed.  In addition to these simulations, an additional 6-week simulation was
performed that considers the 15 seconds following an impact under microgravity conditions
to prepare future investigations closer to the actual gravitational environment on these small
bodies.

\begin{table*}[ht]
\caption[The 5 sets of parameters used for the suite of impact simulations.]{Listing of the 5
  sets of simulation parameters that were used to compare to experimental results
  (see main text for further descriptions of these quantities).  The parameter
  sets (PS$_1$--PS$_5$) were chosen because each provided a good match to the baseline
  experiment, which uses the 90-degree conical projectile impacting the target made up of
  about 45,000 5~mm glass beads.  Each of these parameter sets were used for targets
  comprised
  of the 5~mm beads (5mmPS$_1$--5mmPS$_5$) and targets comprised of the 3~mm beads
  (3mmPS$_1$--3mmPS$_5$), making 10 simulations for each of the 7 projectiles.  Results of
  the simulations are shown in \figtwo{5mm}{3mm}.
  }
\begin{center}
\label{t:impact_params}
\begin{tabular}{cccccccc}
\hline
\hline
Parameter Set&&&$\mu_s$&&$\epsilon_t^*$&&$\varepsilon_n$ \\
\hline
PS$_1$&&&0.000&&1.000&&0.730 \\
PS$_2$&&&0.090&&0.650&&0.950 \\
PS$_3$&&&0.100&&1.000&&0.800 \\
PS$_4$&&&0.180&&0.950&&0.950 \\
PS$_5$&&&0.265&&1.000&&0.950 \\
\hline
\hline
\end{tabular}
\end{center}

\footnotesize{$^*$The quantity $\epsilon_t$ used here is not in fact the true tangential coefficient of restitution,
$\varepsilon_t$, a quantity not straightforward to specify in an SSDEM simulation (see \citealt{Schwartz2012}).  Still,
$\epsilon_t$ has a one-to-one mapping to $C_t$, the coefficient of tangential kinetic friction, and since it can be simpler to work with
dimensionless
quantities of order unity, it is adopted here using the SSDEM stiffness parameter, $k_t$.  Analogous to
$\varepsilon_n$, $\epsilon_t$ is defined as:
\begin{equation*}
  C_t \equiv -2 \ln\epsilon_t \sqrt{\frac{k_t \, {\mu}}{\pi^2 +
      (\ln\epsilon_t)^2}}{\rm .} \label{e:Ct}
\end{equation*}
}
\normalsize
\end{table*}

\begin{figure*}  
  \centering
  \epsscale{0.495}
  \raggedright
  \plotone{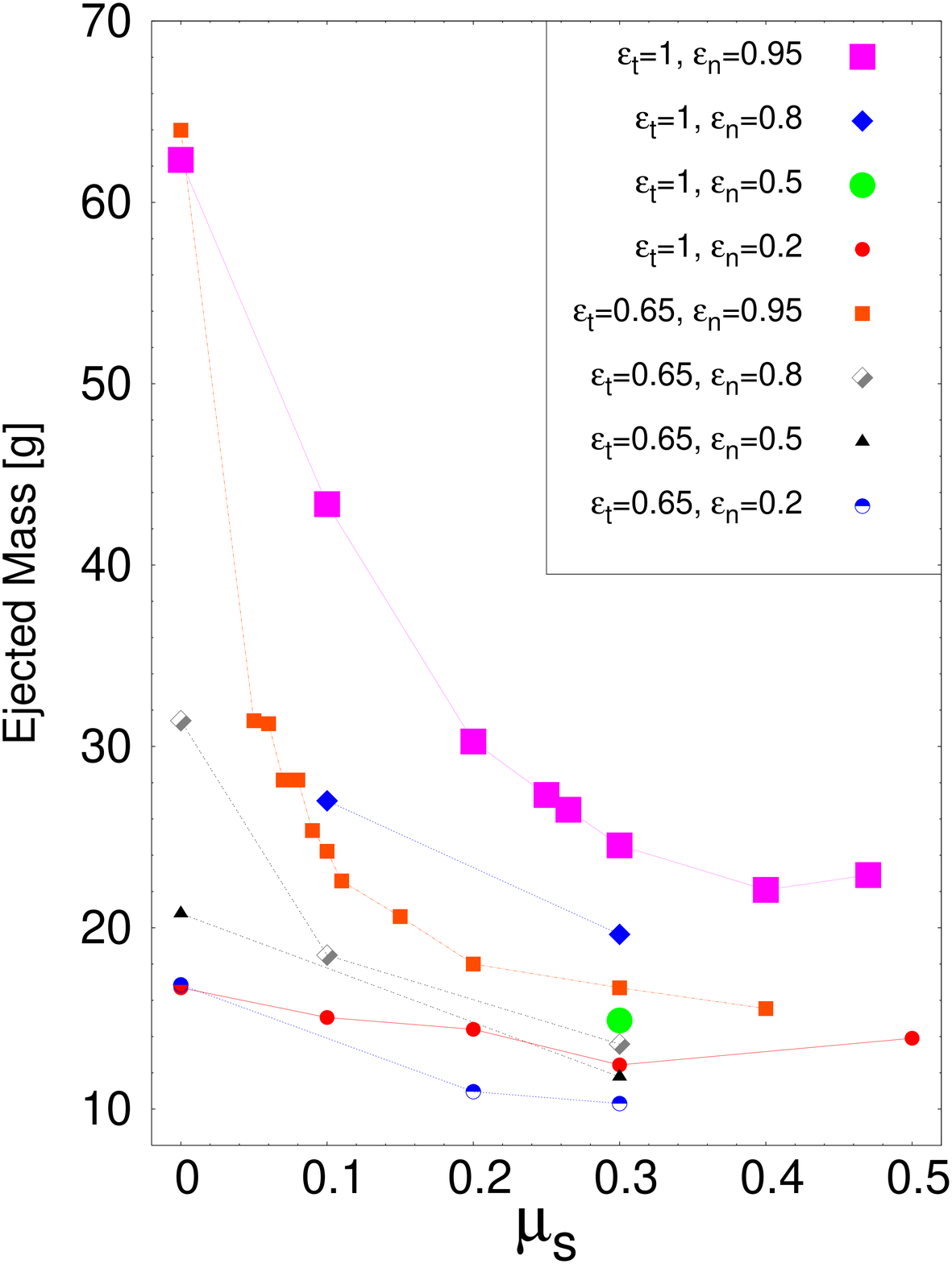}
  \plotone{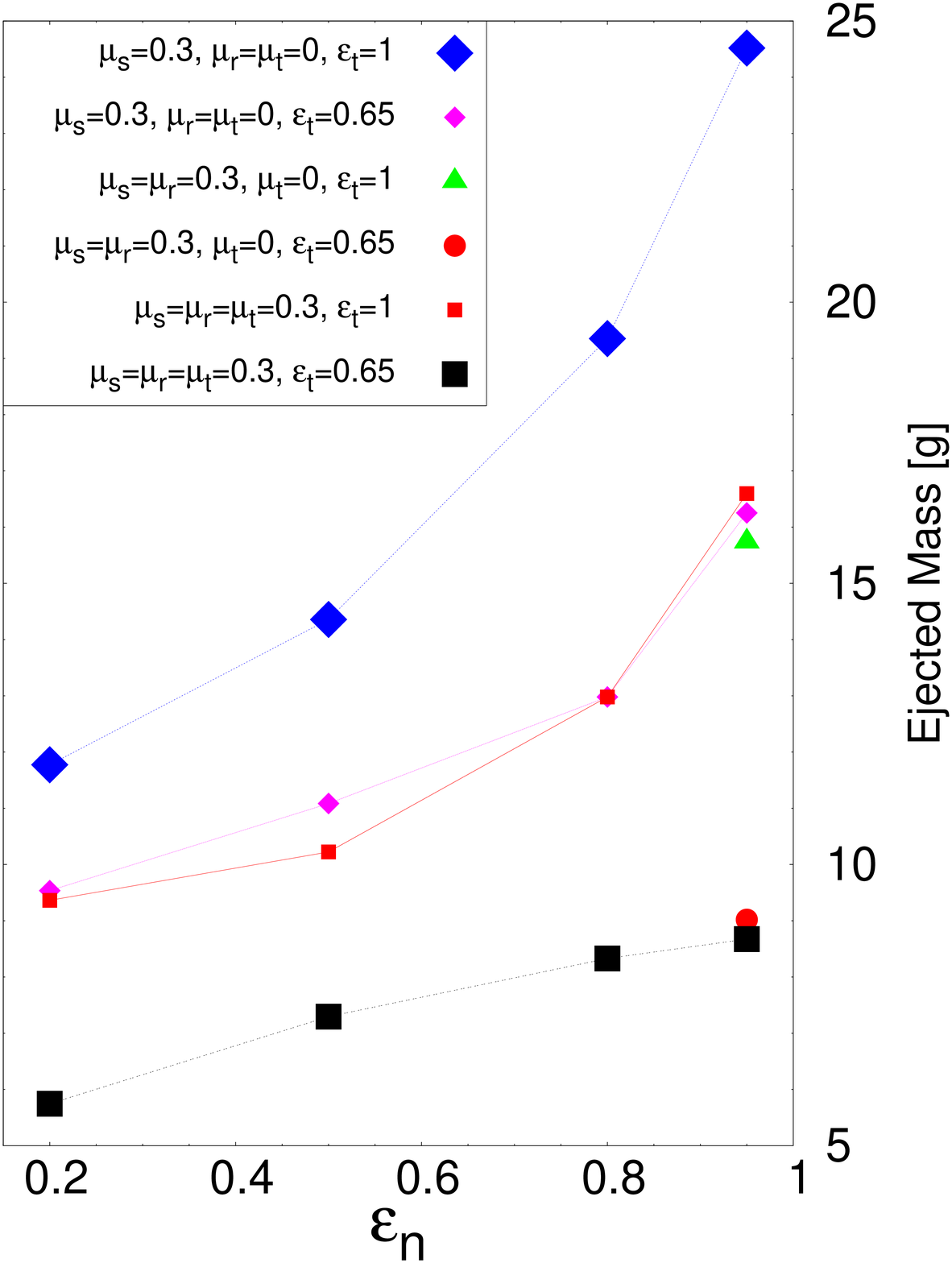}
  \caption[Dependencies of total mass ejected from containers on certain SSDEM paramters]{
  Total mass ejected using the 90-degree cone impactor shape and different SSDEM parameters (see labels on the plots) with selected data taken from the 93-simulation parameter sweep.  Left: total ejected mass plotted against the coefficient of static friction.  Increasing the value of the static friction parameter decreases the amount of mass ejected for low to moderate values of $\mu_s$.  Included are the results for all simulations with $\epsilon_t=1$ and $\epsilon_t=0.65$.  Right: total ejected mass plotted against the coefficient of restitution for cases with $\mu_s=0.3$.  Higher values of the coefficient of restitution parameter tend to increase the amount of mass ejected.  Dependencies of total mass ejected on parameters other than $\mu_s$ and $\varepsilon_n$ can also be inferred (see text).
  }
  \label{f:dMuS}
\end{figure*}

The results from the impact simulations were compared against the experimental results,
and were also analyzed to determine how the values of specific
parameters influence the amount of mass ejected from the container.
Preliminary exploration of the full material-parameter space suggests that $\mu_r$ and $\mu_t$ can affect the penetration depth of the projectile, and thus the total amount of mass ejected, but only when used in conjunction with sufficiently high values of $\mu_s$, typically at values of $\mu_s$ greater than those used in this study.  Used in conjunction with relatively low values of $\mu_s$ ($\leq0.1$; see \tbl{0.1dMuS}), both $\mu_r$ and $\mu_t$ show no discernible effect on the amount of mass ejected.  In simulations with $\mu_s=0.3$, higher values of $\mu_r$ tend to decrease the amount of ejected mass (\fig{dMuS}; right).  The resulting crater shape will also be highly affected by the $\mu_r$ parameter, however crater morphology is deferred to a future study.  Effects on the amounts of mass ejected due to variations in the value of $\mu_t$ were indiscernible from any of these data.
In order to limit the wide parameter space to be explored, and because
they have little influence on the amounts of mass ejected when used in conjunction with the values of the other SSDEM parameters chosen to match the amounts of mass ejected experimentally, the parameters $\mu_r$ and $\mu_t$ were kept at zero within these parameter sets.

\begin{table*}[ht]
\caption[From the 93-simulation parameter sweep, runs using $\mu_s=0.1$ and a range of
  values for $\mu_r$ and $\mu_t$.]{Runs from the 93-simulation parameter sweep that use 
  $\mu_s=0.1$, $\epsilon_t=1$, and $\varepsilon_n=0.2$.  Holding these three parameters
  constant, neither $\mu_r$ nor $\mu_t$ seem to have a discernible effect on the outcome.
  }
\begin{center}
\label{t:0.1dMuS}
\begin{tabular}{ccccccccccc}
\hline
\hline
$\mu_s$&&$\mu_r$&&$\mu_t$&&$\epsilon_t^*$&&$\varepsilon_n$&&Ejected Mass [g] \\
\hline
0.100&&0.000&&0.000&&1.000&&0.200&&15.05 \\
0.100&&0.100&&0.100&&1.000&&0.200&&15.71 \\
0.100&&0.200&&0.200&&1.000&&0.200&&15.38 \\
0.100&&0.300&&0.300&&1.000&&0.200&&15.87 \\
0.100&&0.400&&0.000&&1.000&&0.200&&15.38 \\
0.100&&0.400&&0.200&&1.000&&0.200&&15.54 \\
0.100&&0.400&&0.400&&1.000&&0.200&&15.54 \\
\hline
\hline
\end{tabular}
\end{center}

\footnotesize{$^*$See the equation in the caption of \tbl{impact_params} in regard to the usage of $\epsilon_t$ in this manuscript.
}
\normalsize
\end{table*}

\begin{figure*}  
  \centering
  \epsscale{1}
  \raggedright
  \plotone{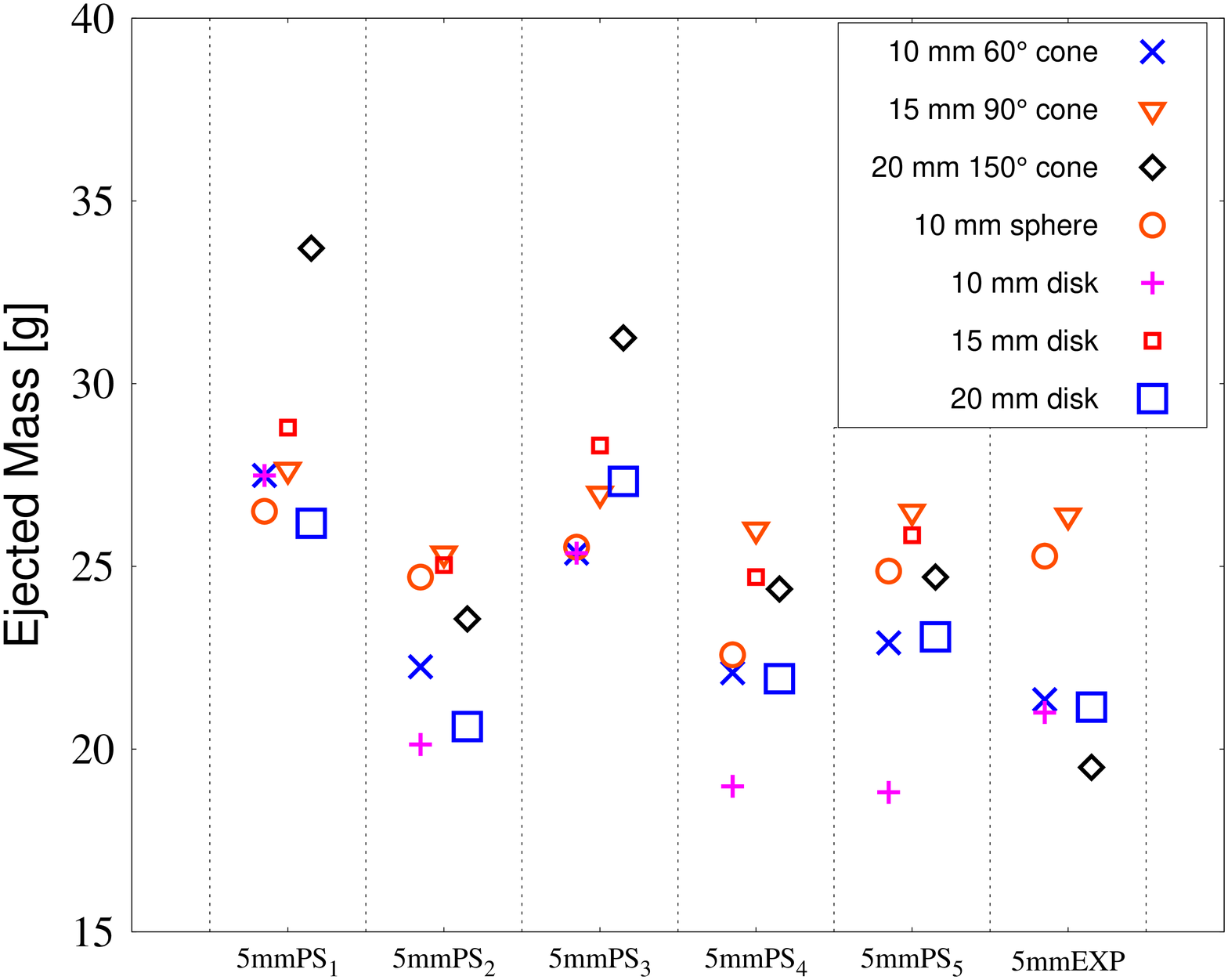}
  \caption[Ejected mass from targets comprised of 5~mm grains]{
  Ejected mass from simulations using 5~mm particle targets, the 5 parameter sets (PS$_1$--PS$_5$; see
  \tbl{impact_params}), and 7 projectiles.
  The experiments that use the 5~mm particle targets (5mmEXP) are also shown for comparison.  (No experiments
  were performed using the 15~mm disk projectile into the target of 5~mm particles.) 
  }
  \label{f:5mm}
\end{figure*}

Using different combinations of 3 parameters, $\mu_s$, $\varepsilon_n$, and $\epsilon_t$, these
5 parameter sets were selected such that they each eject the same amount of mass for one
experimental setup: the 90-degree cone impacting into 5-mm beads, as explained earlier.
Therefore, it must be kept in mind that patterns in the results of the 70-run suite
(\figs{5mm}{3mm}) will be relative to this baseline setup.
Impacts from any projectile shape into either of the 2 targets show that an increase in $\mu_s$ decreases
the amount of mass ejected, a decrease in $\epsilon_t$ (an increase in tangential damping forces; see the
equation in the caption of \tbl{impact_params}) also
decreases the
amount of mass ejected, but that an increase in $\varepsilon_n$ increases the amount of mass ejected.
\Fig{dMuS}, showing data from the 93-simulation parameter sweep, shows that in the ranges of values tested, $\mu_s$ decreases, and that $\varepsilon_n$ and $\epsilon_t$ increase, the amount of total ejected mass.

\begin{figure*}  
  \centering
  \epsscale{1}
  \raggedright
  \plotone{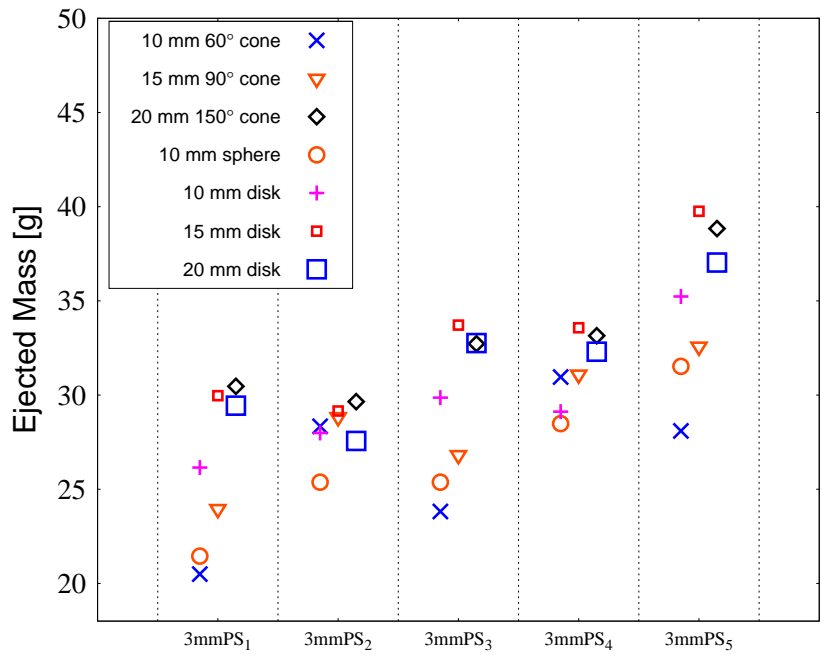}
  \caption[Ejected mass from targets comprised of 3~mm grains]{
  Ejected mass from simulations using 3~mm particle targets, the 5 parameter sets (PS$_1$--PS$_5$; see
  \tbl{impact_params}), and 7 projectiles.
  }
  \label{f:3mm}
\end{figure*}

Results from the use of the 3-mm particles show that the simulations from the 70-run suite with greater static friction tend
to increase mass loss (the parameter sets PS$_1$--PS$_5$ are ordered with increasing
$\mu_s$, while each set has different values for the other parameters).  This result means that $\mu_s$, relative to $\varepsilon_n$ and $\epsilon_t$, has less of an effect
on the total amount of ejected mass for the smaller,
3-mm beads than it does for the larger, 5-mm beads.
(Conversely, if the trend were reversed---that is, a trend showing a decrease in total mass ejected from left to right in \fig{3mm}---this would imply that $\mu_s$ had a greater effect on the amount of mass ejected on the 3-mm beads than it does on the 5-mm beads impacted with the 90-degree cone projectile.)

In the case of the baseline experiment---the 90-degree cone impacting into 5-mm beads---an increase in static friction requires a corresponding increase in $\varepsilon_n$ (PS$_1$, PS$_3$, and PS$_5$) in order to eject the same amount of mass.  In the case of the 3-mm beads (\fig{3mm}), the increases in $\varepsilon_n$ tend to overcompensate for the change in $\mu_s$, showing that the ejection of 3-mm beads is more sensitive to the parameter $\varepsilon_n$ than to the parameter $\mu_s$, relative to the baseline experiment.
At the 5-mm grain size, mass loss does not correlate with static friction across these parameter sets.  This shows that, relative to each other and within the set of projectile shapes used, the simulations involving 5-mm beads tend to have similar dependencies on these two parameters.

Over the 5 parameter sets that use the targets comprised of 5-mm beads (hereafter referred
to as 5mmPS$_1$--5mmPS$_5$), $\varepsilon_n$ seems to be the dominant parameter in
determining mass loss: 5mmPS$_2$, 5mmPS$_4$, 5mmPS$_5$ each share the same high
value of $\varepsilon_n$, namely 0.95.
This means that an increase in $\varepsilon_n$ does less to increase the amount of ejecta for
the simulations with 5-mm beads than for those with 3-mm beads, save for the baseline
experiment and the spherical projectile.
The experimental results using the 5-mm beads appear to match better
the simulations that use an $\varepsilon_n=0.95$ (see \sect{con4} for continued
discussion).
For most projectile shapes, larger values of $\varepsilon_n$ do less to increase mass ejection of
5-mm beads than they do in the baseline simulations.  This is especially the case for the
the 20-mm disk, 150-degree cone, and the 10-mm disk projectiles.
For each simulation using $\varepsilon_n=0.95$ and 5-mm beads, the 90-degree cone projectile
maximizes the amount of ejected mass, which agrees with the experimental results.

\begin{figure*}  
  \centering
  \epsscale{0.19}
  \raggedright
  \plotone{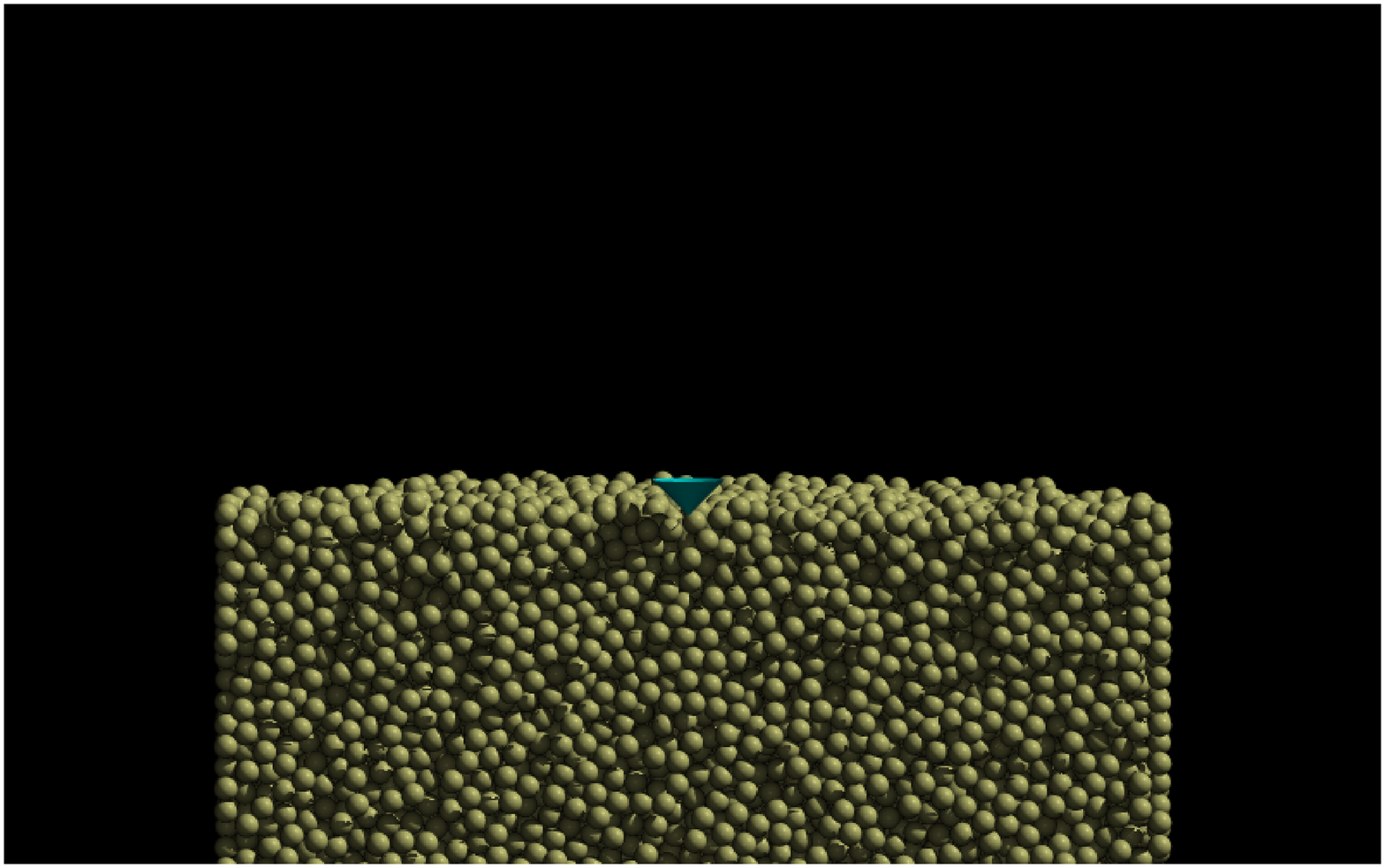}
  \plotone{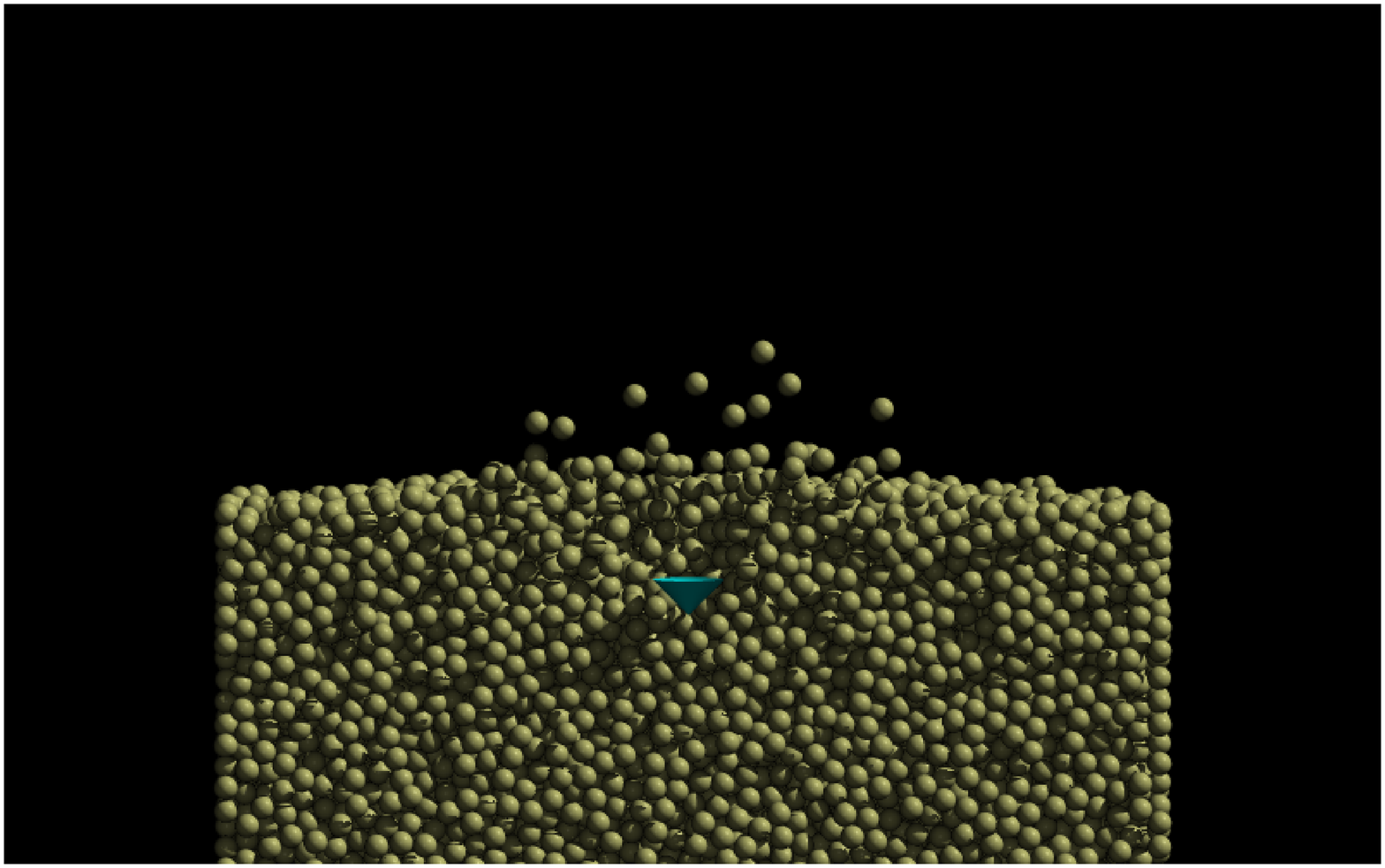}
  \plotone{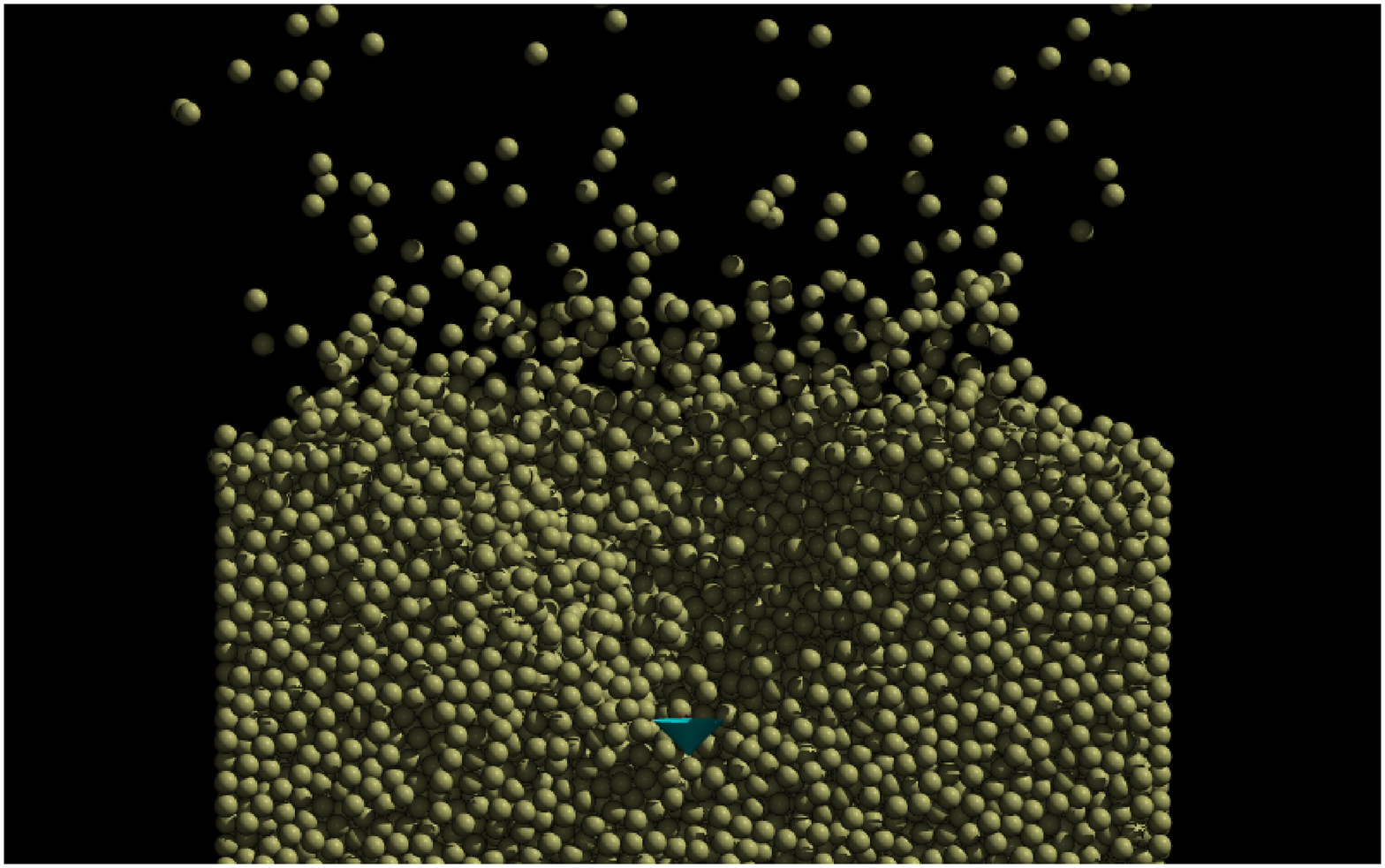}
  \plotone{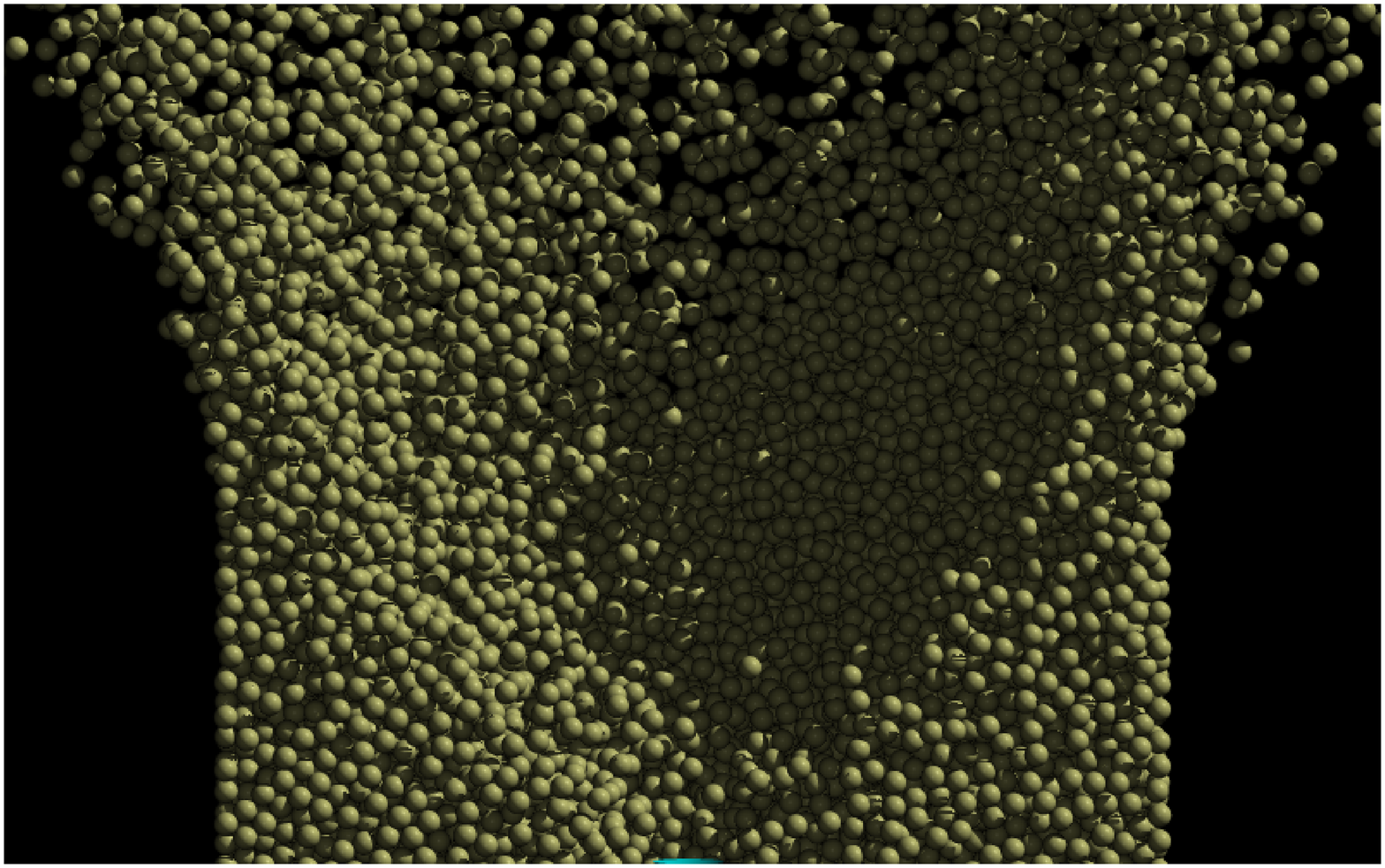}
  \plotone{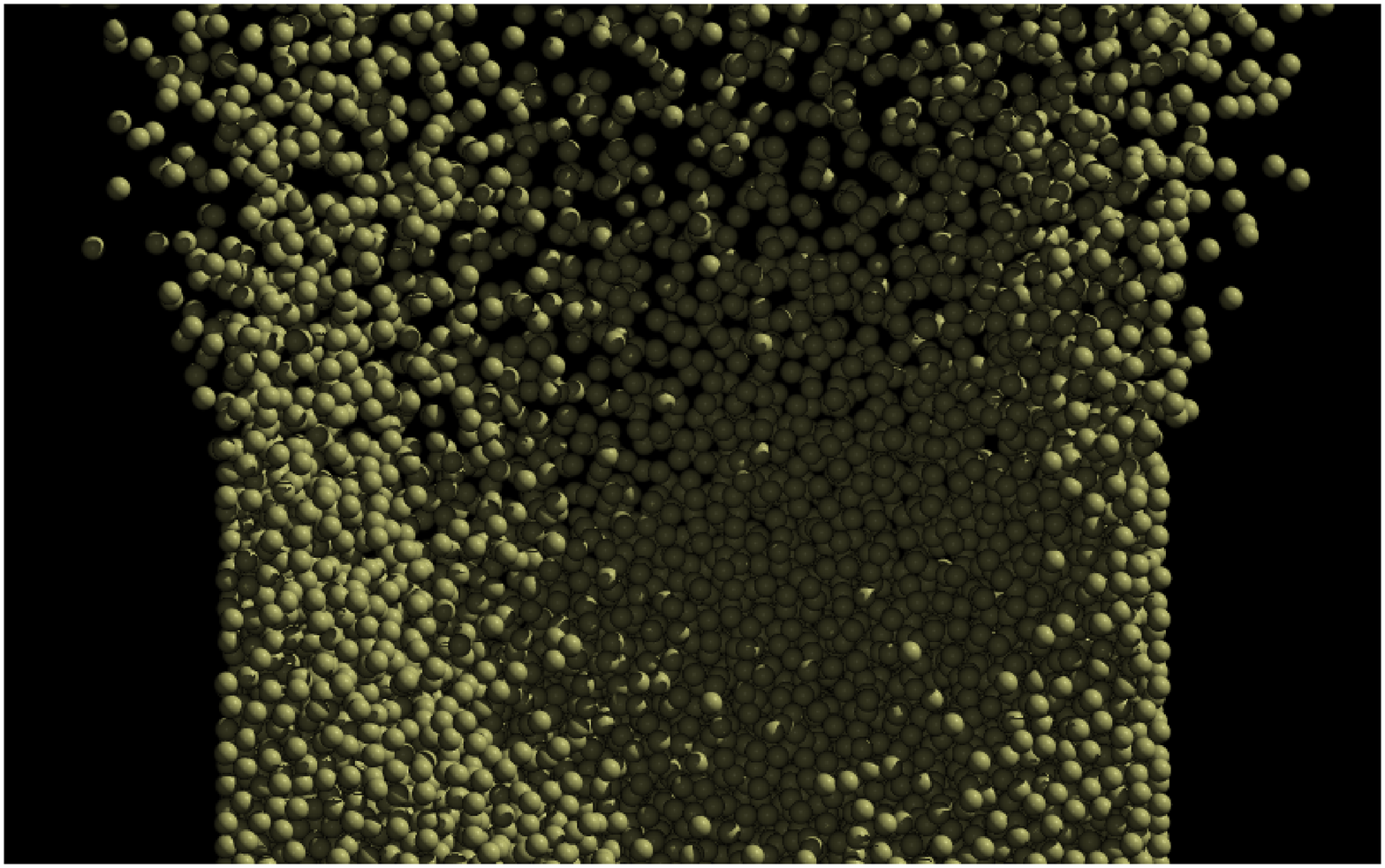}
  \caption[Impact simulation under microgravity conditions]{
  Successive frames showing an 11~m~s$^{-1}$ impact simulation of the 90-degree cone into 3~mm particles under microgravity conditions.  From left to right: time of impact; 10~ms after impact; 100~ms after impact; 1~s after impact; 10~s after impact.  After 10 s, the projectile has descended about 11~cm (just out of the snapshot frame), the ejecta plume near the surface is populated with low-speed ejecta ($<$ 1~cm~s$^{-1}$), and material continues to be expelled slowly from the container.
  }
  \label{f:1ms}
\end{figure*}

\begin{figure*}  
  \centering
  \epsscale{1}
  \raggedright
  \plotone{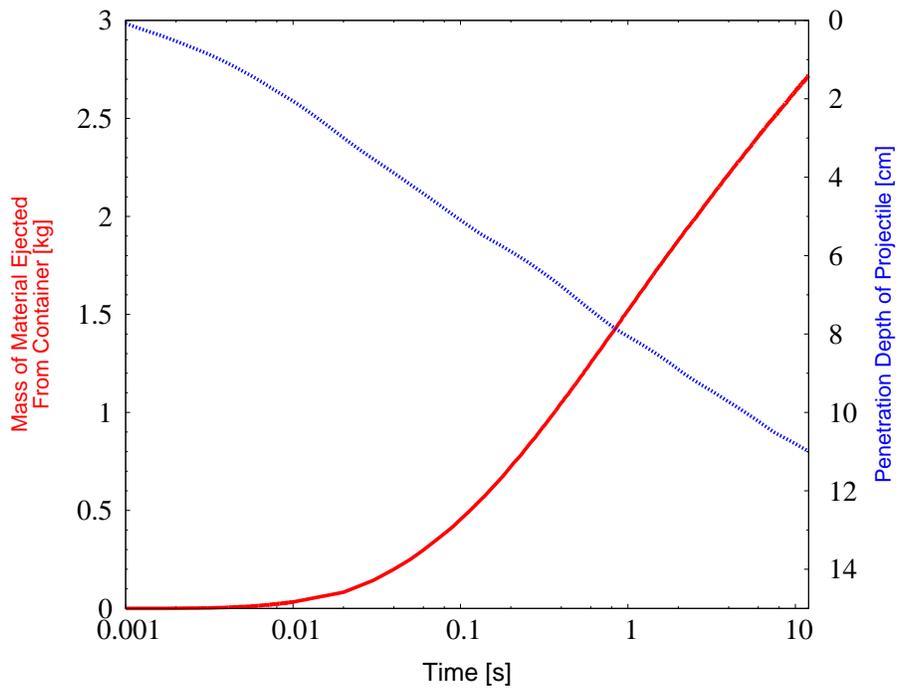}
  \caption[Mass ejected above the rim of the container and projectile penetration depth]{
  Ejected mass (red/solid line) and projectile penetration depth (blue/dotted line) \vs time after impact.  Results obtained from a simulation using 5~mm particle targets with the second
  material parameter set (PS$_2$) under microgravity conditions.  The 90-degree canonical shape was used as the projectile.
  }
  \label{f:microg}
\end{figure*}

One advantage of our numerical approach is that we can place the simulated processes into different
gravitational environments, suitable for different locations throughout the Solar System.  Being primarily
interested in conditions relevant to asteroid sampling mechanisms, outside of this suite of runs under the
influence of Earth's gravity, we also performed a first impact simulation, using the 90-degree canonical
projectile into a 5mmPS$_2$ target under microgravity conditions (10$^{-6}$ Earth gravity).  Snapshots
of the simulation are shown in \fig{1ms}.  Although this simulation was not carried out
as part of a complete study involving different gravitational regimes, which will be the subject of a future
work, it demonstrates the applicability of our approach to the surface environment of a small asteroid.  As
one might expect when using the same impact speed in microgravity environments, 11 m~s$^{-1}$,
boundary conditions (container size) quickly become important.  A
measurement of the projectile penetration depth, and of the amount of ejected material at a given time after
the impact, is given in \fig{microg}.  Because of the longer timescales involved in low-gravity conditions,
simulations need to be run much further forward in time to capture more of the process.  Future studies
will be carried out to obtain a more complete understanding of the nature of these impact processes
under these gravitational regimes.

\section{Discussions, conclusions, and perspectives on this work}
\label{s:con4}

We performed numerical impact simulations to measure the amount
of granular material ejected as a result of impacts from different projectile shapes.
The experiments on which these simulations were based had made use of Hayabusa
spacecraft observations of coarse-grained regolith on Itokawa to determine relevant grain sizes
to use (5~mm) on bodies with similar surfaces.  Targets comprised of 0.5-mm beads were also
used in experiments to explore larger
projectile-to-target grain size ratios.  Firing projectiles of similar mass as the projectile chosen
for the sampling mechanism aboard Hayabusa, the results suggest that the projectile shape
that corresponds to the 90-degree cone was likely to eject the greatest amount of regolith,
especially when using the larger mass ratio.  The simulations performed in this
study did not use 0.5-mm beads for reasons of computational cost (see \sect{sim4}).  In this
study, only numerically modeled containers filled with 5-mm beads and containers filled with
3-mm beads were used.
The limited grain-size regimes of our simulations meant we were unable to explore or account for the systematic decrease seen experimentally in the total amount of ejected mass when particle size is decreased from 5~mm to 0.5~mm.

We developed and used a new feature allowing projectiles to take on non-spherical shapes,
but with the primary limitation that it neglects torques and transverse forces on the projectile.
We explored parameter space based on the experiments that included the 90-degree
cone projectile and the containers filled with 5-mm glass beads.  Ninety-three simulations
were performed to replicate this experiment using a wide range of parameters, holding the precise impact point and target configuration fixed.
Within the range explored, we assessed the effects of 5 different SSDEM parameters (\fig{dMuS}; \tbl{0.1dMuS}).  We found that greater static friction decreases the total amount of mass ejected, although this effect seems to lessen at the larger end of values of static friction explored here (simulations have $0.0\leq\mu_s\leq0.5$).  Also, we found increasing $\varepsilon_n$ and $\epsilon_t$ increases the amount of mass ejected (we did not explore values of $\epsilon_t$ lower than 0.2).  At low values of $\mu_s$ ($\leq0.1$), rolling friction appears to have little if any effect on the amount of mass ejected, however, at higher values of $\mu_s$, an increase in $\mu_r$ decreases the amount of mass ejected.  In the parameter space explored, we did not find $\mu_t$ to affect the amount of mass ejected.

We also explored the effect of varying the precise impact point of the 90-degree cone projectile onto the target and found relatively small 1-$\sigma$ deviations of about 7\% of the mean amount of mass ejected (just under 2~g) and about 3\% of the mean amount of mass ejected (just under 1~g) for the respective cases of the 5-mm beads and the 3-mm beads.

After our preliminary exploration of the material-parameter space using a single projectile, we then
selected 5 parameter sets based on how well the corresponding simulations matched the
experimental results.   These parameter sets, together covering a wide range in parameter
space (omitting $\mu_r$ and $\mu_t$), were then used to simulate impacts for each of the 7 projectiles into the 2
containers (one was filled with 5-mm beads and the other with 3-mm beads).

The simulated impacts
into 5-mm beads matched up well with the experimental results, both in regard to the amount
of mass ejected and to the relative ordering of the projectiles that ejected the most mass (see
\fig{5mm}).  This was especially the case when using beads with a high normal
restitution coefficient of 0.95.
The spherical projectile (Hayabusa-type) in the 5-mm cases shows similar mass loss across
the parameter sets, which means
that these parameters affect the results using the spherical projectile in a similar manner as
they do using the 90-degree cone.  Some of the other projectile shapes, such as the
150-degree cone, showed more variation, and more sensitivity to the value of the normal
coefficient of restitution.
The 90-degree cone projectile ejected a similar amount of mass when impacted into targets
comprised of 3-mm beads as it did into the targets comprised of 5-mm beads, namely in the
${\sim}25$--30-g mass range, but the flat disks and 150-degree cone produced more ejected
mass in many of the cases that used the 3-mm beads.  Since there were no impact
experiments that used 3-mm beads, nor were
there numerical simulations that used 0.5-mm beads, it remains unclear how sensitive the
results of mass ejection are on the projectile-to-target grain size ratios.

For example, could the neglecting of torques felt by the projectile (\sect{method4}) be
responsible for the fact that the
90-degree cone did not eject more mass relative to others in the case of the 3-mm beads?
Neglecting these torques is warranted for the spherically shaped projectile in
this type of impact simulation, however, for shapes that are prone to feel relatively large
torques, this
assumption is less sound.  Although the target is cylindrically
symmetric on average, small asymmetric forces on, say, a thin flat disk, could cause some modest
rotation and lateral motion, influencing the way that energy is delivered to the target grains.  This said,
it should be noted that this did not seem to be a factor in the simulations that used the
spherical projectile and targets of 5-mm beads (\fig{penetrate}).
Additionally, in the experiment, images obtained of a disk projectile penetrating the target
show an extremely symmetrical ejecta plume, which may mean that there is very little rotation or lateral
motion of the projectile during penetration (\fig{500m1}).  Nevertheless, neglecting these degrees-of-freedom on projectiles in simulation could lead to an overestimation of the amount
of material ejected; for these reasons, one should expect this to be more of a factor when considering the
wider disk and the 150-degree cone projectiles.  These projectiles did seem to ``over-perform"
in simulations using 3-mm beads; that is, they seemed to eject more mass relative to the other
projectiles than one might expect based upon the experimental results using other bead sizes.

The 3-mm bead targets were affected more by the combination
of $\varepsilon_n$ and $\epsilon_t$ than they were by $\mu_s$, relative to the baseline simulation.
The 5-mm bead targets seemed to all be affected by the $\mu_s$ parameter in a similar way.
It was found that $\varepsilon_n$ generally affects the targets comprised of 3-mm beads in
a similar way as it affects the baseline simulation.  The targets of 5-mm beads using the other
projectile shapes, with the exception of the sphere, were affected more by the combination of
$\mu_s$ and $\epsilon_t$ than they were by $\varepsilon_n$.
This implies that $\varepsilon_n$ may do more to increase the amount of mass ejected
from targets comprised of smaller grains, and that $\mu_s$ may do more to suppress the
amount of mass ejected from targets comprised of larger grains that are of more comparable
size to the impactor (\ie smaller projectile-to-target grain size ratio).
For these larger target grains, it would seem that surface friction effects play a strong
role, but as target grain sizes
become significantly smaller than the impactor (\ie larger projectile-to-target grain size ratio),
material properties other than the normal restitution coefficient become less important.
This makes some intuitive sense because smaller grains imply a greater number of grains in a given volume, leading to a greater number of collisions, increasing the importance of collisional dissipation.
Although the sample size here is small, these results are worthy of further investigation.

In all, our numerical methodology reproduces important aspects of the experiments, leading to
the same conclusion that the 90-degree cone is the projectile shape that appears to eject the
most 5-mm grain material.  We can therefore have some confidence in applying our approach
in different regimes.  The next steps include the simulation of granular impacts in faster speed
regimes
(ranging from 100~m~s$^{-1}$ to 300~m~s$^{-1}$, which include cases of the Hayabusa-type
sampling mechanism) and simulations under various gravitational regimes, such as
the microgravity environments of asteroids like Itokawa (Hayabusa) and 1999JU$_3$
(Hayabusa2); such a study is underway (see \figtwo{1ms}{microg}).  The integration of impacts
until later times, in order to analyze the resulting crater morphology and ejecta paths, will also
be carried out.  Finally, we note that in low-gravity regimes, and when using small
particles, van der Waals forces can become relevant \citep{London1936,Scheeres2010}, and
so it would be prudent to include
these forces in simulations.  This is a capability of the code that we plan to test.

\section*{ACKNOWLEDGMENTS}

This material is based on work supported by the National Aeronautics and Space
Administration under Grant Nos.\ NNX08AM39G, NNX10AQ01G and NNX12AG29G issued through
the Office of Space Science and by the National Science Foundation under Grant No.\
AST1009579.  S.R.S.~and P.M.\ acknowledge support from the French Space National
Agency (CNES).  S.R.S.\ acknowledges support from the Chateau\-briand
2011 fellowship from the Embassy of France in the United States and partial support from
the European Union Seventh Framework Programme (FP7/2007-2013) under grant agreement
No.\ 282703-NEOShield.
H.Y.Õs contributions
were partially supported by the Japan Society for the Promotion of Science
Grant-in-Aid for Scientific Research on Innovative Areas (Grant No.\ 20200048) entitled
``Innovation of Microgravity Geology."
Most of the computation was performed using the Beowulf
computing cluster (\code{yorp}), run by the Center for Theory and Computation at the
University of Maryland's Department of Astronomy, while many runs were also
performed using that department's public-use machines.  For data visualization, the
authors made use of the freeware, multi-platform ray-tracing package, Persistence
of Vision Raytracer (\code{POV-Ray}).

\bibliographystyle{model2-names}
\bibliography{refs}

\end{document}